\documentstyle[12pt]{article}
\newcommand{\rr}{\rule[-0.2cm]{0cm}{0.8cm}}

\newcommand{\bq}{\begin{equation}}
\newcommand{\ee}{\end{equation}}
\newcommand{\al}{\mbox{$\alpha$}}
\newcommand{\ap}{\mbox{$(\alpha/\pi)$}}

\newcommand{\fr}[2]{\frac{#1}{#2}}

\newcommand{\bi}[1]{\bibitem{#1}}

\newcommand{\hpl}[5]{ \put(#1,#2){\begin{picture}(80,20)
\multiput(#3,0)(#4,0){#5}{\oval(#3,#3)[b]}
\multiput(0,0)(#4,0){#5}{\oval(#3,#3)[t]}  \end{picture}}
}

\newcommand{\hpb}[4]{  \put(#1,#2){\begin{picture}(80,20)
\hpl{0}{0}{4}{8}{3} \hpl{0}{#3}{4}{8}{3} \hpl{0}{#4}{4}{8}{3}
 \end{picture}} }

\newcommand{\hpp}[3]{  \put(#1,#2) {\begin{picture}(80,20)
\hpl{0}{0}{4}{8}{4}    \hpl{0}{#3}{4}{8}{4}
 \end{picture}} }

\newcommand{\vpl}[5]{ \put(#1,#2){\begin{picture}(80,20)
 \multiput(0,#3)(0,#4){#5}{\oval(#3,#3)[l]}
 \multiput(0,0)(0,#4){#5}{\oval(#3,#3)[r]}  \end{picture}}}

\newcommand{\feb}[5]{  \put(#1,#2){\begin{picture}(80,20)
     \thicklines
     \put(0,0){\vector(1,0){#3}}
     \put(#3,#4){\vector(-1,0){#3}}
     \put(#3,0){\vector(0,1){#4}}
     \put(#3,0){\vector(0,1){#5}}   \end{picture}} }

\newcommand{\sei}[2]{
    \put(#1,#2){\begin{picture}(70,70)
             \put(-25,1){\circle{20}}
             \multiput(-16,-8)(8,-5){3}{\oval(8,5)[bl]}
             \multiput(-16,-13)(8,-5){2}{\oval(8,5)[tr]}
             \multiput(-16,10)(8,5){3}{\oval(8,5)[tl]}
             \multiput(-16,15)(8,5){2}{\oval(8,5)[br]}
             \end{picture}}  }

\begin{document}
\large
\baselineskip = 24pt
\vspace{2.0cm}

\begin{center}{\Large \bf Vacuum polarization correction to
the positronium decay rate}\\
\vspace{1.0cm}

{\bf  A.P.Burichenko} \\ Budker Institute of Nuclear
Physics, 630090 Novosibirsk, Russian Federation \\
\vspace{1.0cm}
{\bf  D.Yu.Ivanov} \\
Institute of Mathematics, 630090 Novosibirsk, Russian
Federation \\

\vspace{4.0cm}
{\bf Abstract }\\
\end{center}

Corrections $\sim \al^2 $ to the positronium decay rate,
induced by one--loop vacuum polarization diagram, are
calculated.  Their relative values are $0.4468(3) \ap^2$
for para- and $0.960(3) \ap^2$ for orthopositronium.
\newpage

The measured value of orthopositronium decay rate is
\cite{NGR} \[ \Gamma_{exp}=7.0482(16) \mu s^{-1}. \] At the
same time, its theoretical value (including corrections
$\sim \ap$ and  $\sim \al^{2}\log\al$) is firmly
established \cite{OP,CLS,CL2,Ad,SH,KY} and constitutes \[
\Gamma_{1th}=m\al^{6}\,\fr{2(\pi^{2}-9)}{9\pi}\left[
1-10.28\fr{\al}{\pi}-
\fr{1}{3}\al^{2}\log\fr{1}{\al} \right]=7.03830 \mu s^{-1}. \]

To lift the disagreement between them by
$\ap^2$--corrections, the factor at  $\ap^2$ should be
$250(40)$, which may look unreasonably large and very
unusual for QED.  To remove this discrepancy would be of
importance for clear understanding of the relativistic
bound--state problem in QED.

However, there exists at least two contributions of the
second order with a large coefficient at $\ap^2$. The first
one arises when we square the one--loop correction to the
annihilation amplitude \cite{me}, it is equal to $$
\Gamma_{2th}=28.8(2)(\al/\pi)^{2} \Gamma_0 = 0.00112(1) \mu
s^{-1}.$$ (here $\Gamma_0$ is the lowest--order decay rate
). The second one is the relativistic correction calculated
in \cite{KM} $$ \Gamma_{3th}=46(\al/\pi)^2\Gamma_0=
0.00179(12) \mu s^{-1}. $$ Calculation of the contributions
of two--loop corrections to the annihilation amplitude is
required to obtain the total correction $\sim \al^2$ to the
orthopositronium decay rate.  There exists an opinion
\cite{KMY} that numerical factor at $\ap^2$ in the
amplitude can also be large (actually, the above-mentioned
relativistic correction
\cite{KM} is contained also in that contribution).

  There is a large number of diagrams which contribute  to
the second--order amplitude\footnote{\large If at the
calculation the gauge does not contained infrared
singularities is used, there are 102 diagrams for
orthopositronium and 36 diagrams for parapositronium.}, but
only sum of them is gauge invariant.  Nevertheless, there
is a class of the two--loop diagrams sum of which is gauge
invariant by itself.  It consist of diagrams which contain
photon self--energy insertion ( these diagrams for
orthopositronium (o-Ps) and parapositronium (p-Ps) are
displayed in Fig.2 and 1 respectively).  In this paper we
calculate these corrections both for o-Ps and p-Ps.

To account for the self--energy part in photon propagators,
we substitute
\bq\label{eq:CaLe}
\fr{g^{\mu \nu}}{k^2+i\epsilon} \; \rightarrow \;
 \fr{g^{\mu \nu} }{k^2+i\epsilon} \; + \; \int _{4m^2}
^\infty \; R(s)  \; \fr{g^{\mu \nu} }{k^2 - s +i\epsilon} d
s  \; \; \; , \ee (K\"{a}llen --- Lehmann representation);
to first order in $\al$,
\bq\label{eq:SP}
R(s) \; = \; \fr{\al}{3 \pi}  \; \; \sqrt{ \fr{s-4m^2}{s} }
\; \;
\fr{s+2m^2}{s^2} \; \; \;. \ee
Hence the considered contribution to positronium  decay
rate is of the form
\[\delta \Gamma = \int _{4m^2} ^\infty \; R(s)  \; Q(s) \;
d s \; \; \;, \]
where $Q(s)$ is the one--loop correction to the decay rate
calculated with the massive virtual photon $(m_\gamma =
\sqrt{s} \;)$.

The calculation of corrections $\sim \al^2$  that are
induced by vacuum polarization, was performed in the same
manner as calculation of corrections $\sim \al$, which is
described in the previous papers
\cite{CLS,Ad,me}  .
Our calculation differs from the one-loop one by the
substitution (\ref{eq:CaLe}) , in which we must deal only
with second term  in the right--hand side, that introduced
a one more integration compared to one--loop case.
Appearing integrals were found numerically using convenient
substitution
\mbox{$s=4m^2/(1-x^2)$} .
To sufficient accuracy the initial particles may be
considered being at rest.  We use covariant summation over
polarizations of the final photons.  The results of
computing of various diagrams are  presented in Tables 1,2.
Thus, the total corrections to the positronium decay rate,
induced by vacuum polarization, constitute
\[\delta \Gamma_{4th}^{p-Ps} = 0.4468(3) \;  \ap^2 \;
\Gamma^{p-Ps}_0= 0.0194 \mu s^{-1} \ , \]
\[\delta \Gamma_{4th}^{o-Ps} = 0.960(3) \; \ap^2 \;
\Gamma^{o-Ps}_0= 3.735(12)\cdot 10^{-5} \mu s^{-1} \ . \]

Let us note  that the main part of the complete result
originates as contribution of the diagram $2(f)$, which
equals to \mbox{$-\fr{8}{9} \ap \Gamma^1_a$}  ( here
\mbox{$\Gamma^1_a$ } is part of $\al$-correction, which
arises due to graph  shown  in Fig.3).

In order to check the results of the numerical calculation
we also computed differential decay rate of o-Ps using
three-dimensionally transverse summation over the final
photons polarizations. The obtained  result coincides with
the differential width which was calculated employing  the
covariant photon polarization sums ( though contributions
of individual graphs do depend on manner of the summation,
with the exception of the  contribution of 2(f), which is
gauge invariant by itself).

Another method to check our results is as follows.  Let
$T_i(s)$ to be the contribution to  the  $Q(s)$  due  to
the diagram $(i)$ ($i=a,b,c$ for p-Ps and $i=a,..,f$ for
o-Ps ).  Their asimptotics at $s\gg1$ are $T_i(s)=A_i/s$.
The values of the $A_i$'s were calculated analytically for
all  contributions  to the p-Ps width  and for $i=a,b,c,d$
in the case of o-Ps.  We examined that the numerically
obtained $T_i$'s have the asimptotics which coincide with
those found analytically.  In the case of o-Ps the check
was made for the differential width.  In addition, for the
graph $2(f)$ $A_f$ may be related to the part of correction
$\sim \al$ to the o-Ps decay rate, which arises due to
graph 3; $\; \;$ $A_f$ which was found in this way is in
agreement with the   numerical results.

If we add obtained in that paper contribution to the
orthopositronium decay rate $\Gamma^{o-Ps}_{4th}$ to ones
calculated earlier $\Gamma_{1th}$, $\Gamma_{2th}$,
$\Gamma_{3th}$, it is obtained the result which still too
different with the experimental one. It is conceivable that
further two loops calculations will result in the better
accordance between theory and experiment.

One of the authors (A.B.) wish to thank I.B.Khriplovich for
useful discussions.

\newpage
\begin{center}
Table 1.   Various contributions to p-Ps decay rate.
\end{center}
{\normalsize
    \begin{tabular}{||c|c|c|c||} \hline
  $a$ & $b$ & $c$ & total  \\ \hline
\rr  0.0158(1)  & 0.0975(1)  &0.3335(1)  & 0.4468(3) \\
 \hline
\end{tabular}}

\begin{center}
Table 2.   Various contributions to o-Ps decay rate.
\end{center}
{\normalsize
    \begin{tabular}{||c|c|c|c|c|c|c||} \hline
  $a$ & $b$ & $c$ & $d$ & $e$ & $f$& total  \\
  \hline
\rr  0.151(1)  &-0.0895(3)  &0.210(1)  &0.062(1) &
-0.094(1) & 0.720(1) &
  0.960(3) \\ \hline
\end{tabular}}


\large

\newpage
\begin{figure}
\begin{picture}(460,540)

  \put(0,400){\begin{picture}(460,140)
            \put(50,0){\begin{picture}(120,140)
              \feb{0}{0}{70}{80}{80}
              \hpp{72}{0}{80}
              \vpl{35}{2}{4}{8}{4}
              \vpl{35}{50}{4}{8}{4}
              \put(35,40){\circle{16}}
              \put(50,-30){(a)}
            \end{picture}}
            \put(280,0){\begin{picture}(200,140)
              \feb{0}{0}{70}{80}{80}
              \hpp{72}{0}{80}
              \sei{70}{40}
              \put(40,-30){(c)}
            \end{picture}}
 \end{picture}}

  \put(0,220){\begin{picture}(460,240)

           \put(100,0){\begin{picture}(100,140)
              \feb{0}{0}{70}{70}{70}
              \hpp{72}{0}{70}
              \put(50,20){\circle{16}}
              \multiput(64,27)(6,4){2}{\oval(8,4)[tl]}
              \multiput(56,27)(6,4){2}{\oval(8,4)[br]}
              \multiput(38,0)(4,6){2}{\oval(4,8)[tl]}
              \multiput(38,8)(4,6){2}{\oval(4,8)[br]}
           \end{picture}}

           \put(230,0){\begin{picture}(100,140)
              \feb{0}{0}{70}{70}{70}
              \hpp{72}{0}{70}
              \put(53,50){\circle{16}}
              \multiput(64,41)(6,-4){2}{\oval(8,4)[bl]}
              \multiput(56,41)(6,-4){2}{\oval(8,4)[tr]}
              \multiput(41,70)(4,-6){2}{\oval(4,8)[bl]}
              \multiput(41,62)(4,-6){2}{\oval(4,8)[tr]}
           \end{picture}}

             \put(200,-30){(b)}
    \put(170,-130){\large Fig. 1 }
   \end{picture}}

\end{picture}
\end{figure}

\newpage
\clearpage
\newpage
\begin{figure}
\begin{picture}(460,540)

 \put(20,360){\begin{picture}(460,140)
       \put(0,70){\begin{picture}(120,140)
             \vpl{30}{2}{4}{8}{4}
             \vpl{30}{50}{4}{8}{4}
             \put(30,40){\circle{16}}
             \hpb{62}{0}{40}{80}
             \feb{0}{0}{60}{80}{40}
             \put(50,-30){(a)}
             \end{picture}}

       \put(220,70){\begin{picture}(200,140)
          \put(0,0){\begin{picture}(100,140)
               \hpb{67}{0}{58}{96}
               \feb{0}{0}{65}{96}{48}
               \sei{65}{29}
          \end{picture}}
           \put(115,0){\begin{picture}(100,140)
             \hpb{67}{0}{33}{96}
             \feb{0}{0}{65}{96}{30}
               \sei{65}{61}
             \end{picture}}
             \put(100,-30){(b)}
             \end{picture}}
  \end{picture}}

 \put(20,260){\begin{picture}(460,140)

      \put(340,0){\begin{picture}(100,140)
             \feb{0}{0}{65}{80}{80}
             \hpb{67}{0}{40}{80}
               \sei{65}{40}
             \put(50,-30){(d)}
             \end{picture}}

       \put(0,0){\begin{picture}(200,140)
          \put(0,0){\begin{picture}(100,140)
             \feb{0}{0}{65}{80}{50}
             \hpb{67}{0}{55}{80}
             \put(55,0){\begin{picture}(100,100)
               \put(-10,20){\circle{16}}
               \multiput(4,27)(6,4){2}{\oval(8,4)[tl]}
               \multiput(-4,27)(6,4){2}{\oval(8,4)[br]}
               \multiput(-22,0)(4,6){2}{\oval(4,8)[tl]}
               \multiput(-22,8)(4,6){2}{\oval(4,8)[br]}
               \end{picture}}
          \end{picture}}

   \put(130,0){\begin{picture}(100,140)
               \feb{0}{0}{65}{80}{25}
               \hpb{67}{0}{30}{80}
      \put(15,0){\begin{picture}(100,100)
                 \put(30,58){\circle{16}}
                 \multiput(44,51)(6,-4){2}{\oval(8,4)[bl]}
                 \multiput(36,51)(6,-4){2}{\oval(8,4)[tr]}
                 \multiput(21,80)(4,-6){2}{\oval(4,8)[bl]}
                 \multiput(21,72)(4,-6){2}{\oval(4,8)[tr]}
                 \end{picture}}
               \end{picture}}
               \put(130,-30){(c)}
           \end{picture}}

   \end{picture}}

  \put(40,100){\begin{picture}(460,140)
       \put(260,0){\begin{picture}(170,140)
             \hpl{37}{35}{4}{8}{2}
             \put(63,35){\circle{24}}
             \hpl{77}{35}{4}{8}{2}
             \hpl{133}{35}{4}{8}{3}
             \hpl{123}{18}{4}{8}{4}
             \hpl{123}{52}{4}{8}{4}
             \put(111,35){\circle{40}}
             \thicklines
             \put(35,35){\vector(-1,1){35}}
             \put(0,0){\vector(1,1){35}}
             \put(50,-30){(f)}
             \end{picture}}

   \put(-20,0){\begin{picture}(200,140)
           \put(0,0){\begin{picture}(100,140)
               \hpb{60}{0}{72}{43}
               \feb{0}{0}{57}{72}{42}
               \put(35,48){\circle{12}}
               \multiput(17,72)(6,-6){3}{\oval(6,6)[bl]}
               \multiput(17,66)(6,-6){3}{\oval(6,6)[tr]}
               \multiput(45,40)(6,-8){3}{\oval(6,8)[bl]}
               \multiput(39,40)(6,-8){3}{\oval(6,8)[tr]}
            \end{picture}}

           \put(115,0){\begin{picture}(100,140)
               \hpb{60}{0}{72}{33}
               \feb{0}{0}{57}{72}{30}
             \put(35,24){\circle{12}}
             \multiput(17,0)(6,6){3}{\oval(6,6)[tl]}
             \multiput(17,6)(6,6){3}{\oval(6,6)[br]}
             \multiput(45,32)(6,8){3}{\oval(6,8)[tl]}
             \multiput(39,32)(6,8){3}{\oval(6,8)[br]}
             \end{picture}}
             \put(110,-30){(e)}
          \end{picture}}

    \put(150,-100){\large Fig. 2 }

   \end{picture}}

\end{picture}
\end{figure}

\newpage
\clearpage
\newpage
\begin{figure}
\begin{picture}(460,540)

  \put(180,300){\begin{picture}(320,140)
             \hpl{37}{35}{4}{8}{4}
             \hpl{109}{35}{4}{8}{3}
             \hpl{99}{18}{4}{8}{4}
             \hpl{99}{52}{4}{8}{4}
             \put(87,35){\circle{40}}
             \thicklines
             \put(35,35){\vector(-1,1){35}}
             \put(0,0){\vector(1,1){35}}
    \put(20,-80){\large Fig. 3 }
   \end{picture}}

\end{picture}
\end{figure}

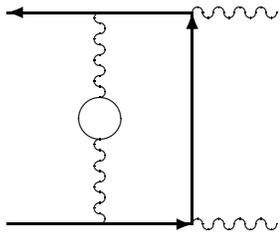
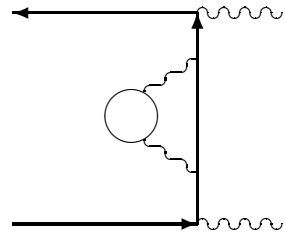
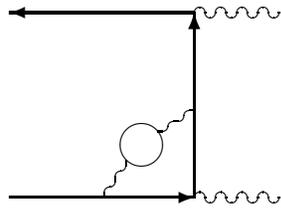
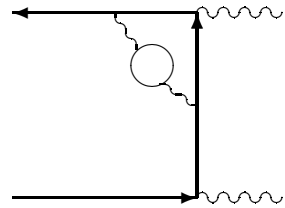
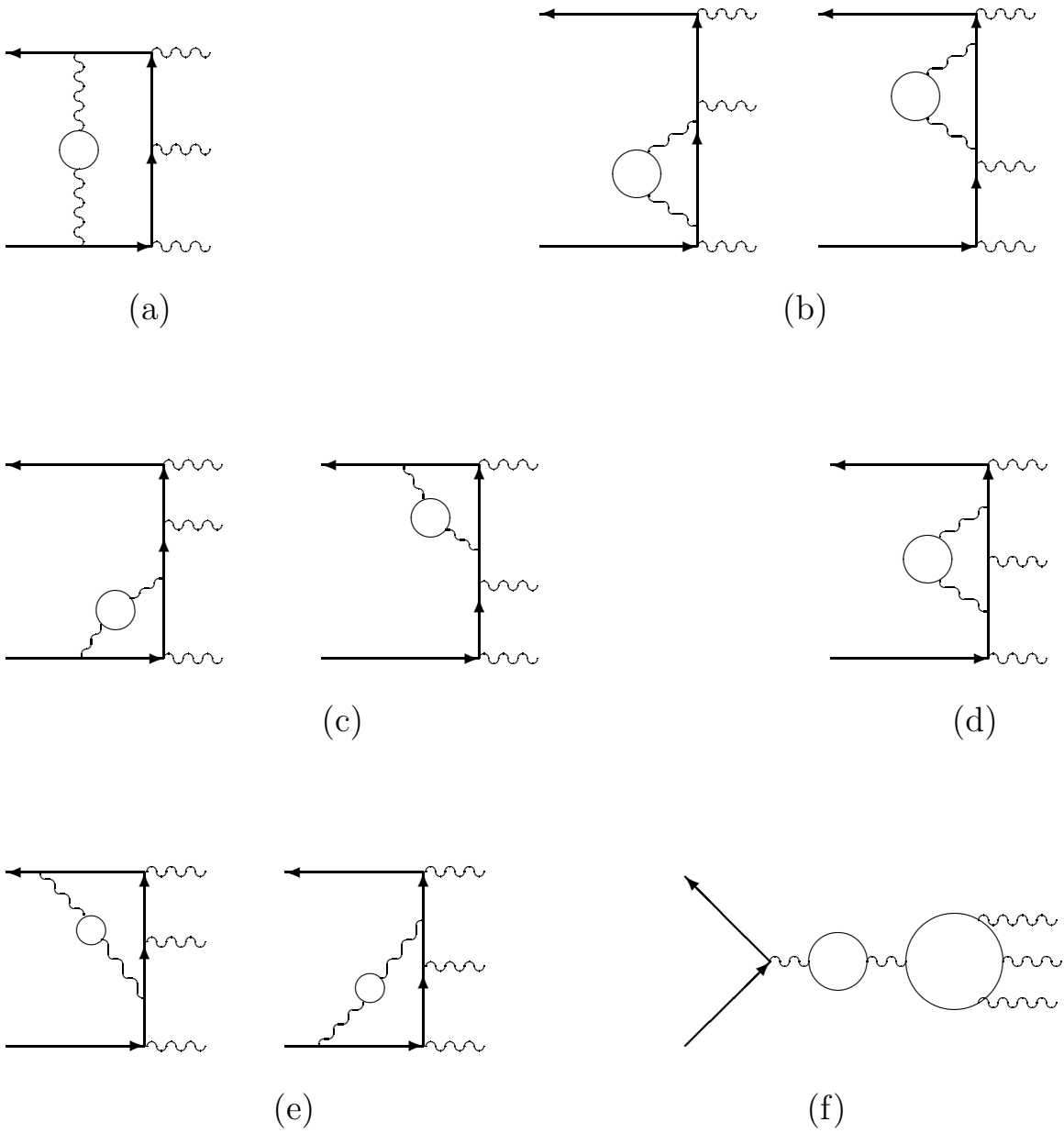
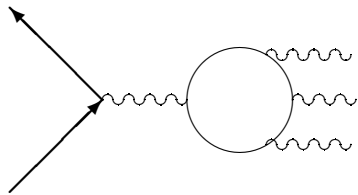
\end{document}